\documentclass[preprint,aps,prd,12pt,preprintnumbers,eqsecnum,nofootinbib]{revtex4-2}
\usepackage{xcolor}
\usepackage{braket}
\usepackage{amssymb,amsmath,amsfonts,comment,latexsym,graphicx,epstopdf,float}
\usepackage{color}
\usepackage{fancyvrb}
\usepackage{chngcntr}
\usepackage{etoolbox}
\usepackage{ulem}
\usepackage{array}
\usepackage{bbold}
\usepackage{subfigure}
\usepackage{slashed}
\usepackage{multirow}

\usepackage[colorlinks=true,citecolor=blue,linkcolor=red,filecolor=cyan,urlcolor=magenta]{hyperref}

\usepackage{float}

\newcommand{\be}{\begin{equation}}
\newcommand{\ee}{\end{equation}}
\newcommand{\ba}{\begin{eqnarray}}
\newcommand{\ea}{\end{eqnarray}}

\newcommand{\grts}{\raise.3ex\hbox{$>$\kern-.75em\lower1ex\hbox{$\sim$}}}
\newcommand{\lets}{\raise.3ex\hbox{$<$\kern-.75em\lower1ex\hbox{$\sim$}}}

{\catcode`\|=\active\gdef\Braket#1{\left<\mathcode`\|"8000\let|\bravert 
{#1}\right>}}

\def\bravert{\egroup\,\vrule\,\bgroup}

\usepackage{color}
\usepackage{amssymb,amsmath,amsfonts}
\usepackage{epstopdf} 
\usepackage{braket}

\usepackage{autobreak}

\begin{document}
%
%
\title{\vspace*{0.5in} 
Tuning towards the edge of a dark abyss:  Implications of a tuning
    paradigm on the hierarchy between the weak and dark matter scales
\vskip 0.1in}
\author{Christopher D. Carone}\email[]{cdcaro@wm.edu}
\author{Noah L. Donald}\email[]{nldonald@wm.edu}
\affiliation{High Energy Theory Group, Department of Physics,
William \& Mary, Williamsburg, VA 23187-8795, USA} 
%
\date{February 3, 2025}
%
%
\begin{abstract}
It has recently been suggested that tuning towards the boundary of the positivity domain of the scalar potential may explain the separation
between the electroweak scale and the unification scale in a grand unified theory.  Here we explore the possibility that the same type of tuning 
might  account for the generation of the electroweak scale from a much lighter dynamically generated scale in a dark sector.  We present a 
model that realizes this idea and provides a proof of principle that the same dark sector can include a 
viable dark matter candidate.
\end{abstract}

\maketitle
\newpage
\section{Introduction}
Over the past 30 years, there have been a vast number of proposals for physics beyond the standard model (BSM) that have been motivated by 
naturalness.  Quantum corrections to the Higgs boson squared mass are quadratically divergent; the bare mass in the Lagrangian must be tuned if the 
electroweak scale is to remain far below the Planck scale or any other high physical scale that may exist in nature.  The minimal supersymmetric 
standard model and its many variations are the best known class of theories that address this problem~\cite{Martin:1997ns}.  Little Higgs 
models~\cite{Perelstein:2005ka} and various higher-derivative extensions of the standard model~\cite{Grinstein:2007mp,Carone:2008iw}   have 
also been put forward with the same goal in mind.  

At present, there is no experimental evidence in favor of any BSM model with new particles or interactions designed to cancel the quadratic 
divergences of the standard model~\cite{PDGNavas}.  While the scale at which such cancelations occur can be raised beyond the increasingly stringent 
lower bounds from collider experiments, the potential link between this scale and the scale of electroweak symmetry breaking is gradually lost.    
For this reason, the non-observation of BSM physics at the Large Hadron Collider (LHC) makes models of this type less compelling and motivates the 
consideration of alternative scenarios.

One possibility is that the tuning of Lagrangian parameters is something that simply occurs in nature.  Rather than 
focusing on eliminating the tuning,  it may be productive as a first step to consider what kinds of tunings may be favored or characterized by 
simple principles.  This may ultimately help with understanding the origin of the tuning, whether it is related to dynamics, anthropic selection 
or some other mechanism.  A recent proposal regarding tuning ``to the edge of the abyss"~\cite{Georgi:2024rxw} is motivated in this way
and is the topic of further study in the present work.    

The edge of the abyss in Ref.~\cite{Georgi:2024rxw} refers to the boundary of the positivity domain of the potential, that is, the parameter 
values for which the potential is on the verge of turning over and becoming unbounded from below.   Near this boundary, there are directions in which the potential is nearly flat.  Introduction of a cubic term with a dimensionful coupling defined by a scale, say $M_0$, can tilt the potential leading to a vacuum expectation value (vev) $v$ that is much larger than $M_0$: $v \gg M_0$.  The generation of a large energy scale from a small one goes against the usual assumptions of effective field theory; Ref.~\cite{Georgi:2024rxw} showed that this separation is consistently maintained when loop corrections are included in the potential, and suggested that the hierarchy between the electroweak scale and the scale of grand unification might arise by such a tuning.

Other approaches to the naturalness problem that challenge the expectations of effective field theory have appeared in the literature previously.  For example, Froggatt and Nielsen have argued that the top quark and Higgs masses may be understood from the principle that there are exactly degenerate vacua at the weak and Planck scales, a critical point where there can be a coexistence of phases~\cite{Froggatt:1995rt}.   More recently, there have been proposals that explain the smallness of the Higgs boson mass via self-organized criticality~\cite{criticality}, a dynamical mechanism that drives a system towards a critical point~\cite{Bak:1987xua}, in this case the vanishing value of the Higgs boson mass, which separates the phases of broken and unbroken electroweak symmetry.  While these proposals on near-criticality and the abyss scenario involve special points where the vacuum is on the verge of becoming unstable, the abyss scenario also involves an instability that does not correspond to a phase boundary, namely the point where the classical potential becomes unbounded from below; the symmetry breaking phases are controlled by an independent model parameter.  The well-known metastability of the Higgs potential in the standard model~\cite{Isidori:2001bm} might give some motivation for considering beyond-the-standard-model scenarios that live relatively close to instability. Whether the assumed form of the potential in the abyss scenario can be understood via some dynamical mechanism is an interesting question; here we focus instead on a new phenomenological application. 

The abyss scenario of Ref.~\cite{Georgi:2024rxw} explored how the electroweak scale may be related to a much higher energy scale, while  here we consider the possibility that a similar tuning may relate the electroweak scale to a dark sector that is characterized by a much lower energy scale.  Instead of a cubic term to tilt the potential, strong dynamics in the dark sector generates a linear term in the scalar potential 
(which, of course, could be subsequently shifted away to generate cubic terms).  Tuning to the edge of the positivity domain of this potential (a boundary
that we refer to as the edge of the ``dark abyss") separates the dark and the electroweak scales, where the latter is triggered by scalar mixing.  This construction leads to a scenario that is significantly constrained once the Higgs vev and mass are fixed~\cite{ParticleDataGroup:2024cfk}, and a mixing angle in the scalar sector is restricted to be below its experimental bound~\cite{Claude:2021sye}. 

Our paper is organized as follows.  In the next section, we define our model and the tuning that leads to separation of the dark and electroweak energy scales.  In Sec.~\ref{sec:pheno} we discuss the phenomenology of the dark sector and show, for the purposes of illustration, how a viable dark matter candidate can be obtained.  We also point out other similar dark sectors to which the present approach might be applied.  In Sec.~\ref{sec:conc}, we summarize our conclusions.

\section{The Model}
We work with a model of dynamical dark chiral symmetry breaking proposed by one of the authors and Ramos in Ref.~\cite{Carone:2015jra}.  The gauge group of the model is $G_{\rm SM} \times$ SU(N)$_{D} \times $SU(2)$_{D}$ where $G_{\rm SM}$ is the gauge group of the standard model, and SU(N)$_D$ is an unbroken ``dark color" group that becomes strongly coupled in the infrared.  The dark sector matter includes a left-handed doublet and two right-handed singlets under the SU(2)$_D$ gauge group. 
\begin{equation} 
\Upsilon_L = \left(\begin{array}{c} p_L \\ m_L \end{array} \right)\, , \,\,\,\,\, p_R\, , \,\,\,\,\,m_R \,\,\,,
\label{eq:darkfermions}
\end{equation} 
which transform in the fundamental representation of the dark color group.  Given this fermion content, we take $N$ to be even so that the theory is free of the SU(2) Witten anomaly~\cite{Witten:1982fp}. The 
dark sector also includes an SU(2)$_D$ doublet field $\phi$ that spontaneously breaks the SU(2)$_D$ gauge group.   
We call the fields $p$ and $m$ in Eq.~(\ref{eq:darkfermions}) for the following reason:  We 
impose a $Z_3$ symmetry under which the dark sector fields 
transform as
\begin{equation}
\begin{array}{ccccccc}
\Upsilon_{L} \rightarrow{} \Upsilon_{L}, &\hspace{1em} & p_{R} \rightarrow{} \omega \, p_{R}, & \hspace{1em} & m_{R} \rightarrow{} \omega^{2} \, m_{R}, & \hspace{1em} & \phi \rightarrow{} \omega \, \phi \\
\end{array}
\label{eq:z3y}
\end{equation} 
where $\omega^{3}=1$.  This can be thought of as a subgroup of a fictitious ``dark hypercharge" gauge symmetry, U(1)$_D$,  under 
which the $\Upsilon$ 
doublet is neutral and the $p$ ($m$) fields have charges $+1/2$ ($-1/2$).   The vev of $\phi$ would 
break SU(2)$_D \times$U(1)$_D$ down to an unbroken U(1), under which the $p$ and $m$ fields (of either chirality) would have charges 
$+1/2$ and $-1/2$ respectively.   (Note the similarity to the structure of the technicolor model in Ref.~\cite{Carone:1993xc}).  In our case, 
we have no continuous dark hypercharge symmetry, and the unbroken symmetry is a $Z_3$ which is the diagonal 
subgroup of a $Z_3$ generated by the diagonal generator of SU(2)$_D$ and the $Z_3$ defined in Eq.~(\ref{eq:z3y}).  Thus, the present model
is economical in that it has no massless dark photon; nevertheless, the dark fermions are charged under this residual $Z_3$, with the Dirac fermions transforming as $p \rightarrow \omega \, p$ and $m \rightarrow \omega^2 \,m$.   Since the residual $Z_3$ symmetry can be embedded into a continuous gauge symmetry, it meets the requirements of a discrete gauge symmetry~\cite{Ibanez:1991hv}; such symmetries may be defined {\it ab initio}, {\it i.e.}, without an explicit embedding, and are thought not to be spoiled by quantum gravitational effects~\cite{Banks:1991xj}.   This makes the symmetry useful for 
stabilizing a dark matter candidate, as we discuss later.

Taking into account this symmetry structure, the allowed dark sector Yukawa couplings are
\begin{equation}
-\mathcal{L}_{y} = y_{+}\overline{\Upsilon}_{L}\Tilde{\phi} \, p_{R} + y_{-}\overline{\Upsilon}_{L}\phi \, m_{R} + h.c. ,
\end{equation}
where $\Tilde{\phi} \equiv i \sigma^{2}\phi^{*}$.  Letting $\Upsilon_{R} = (p_{R},m_{R})$ and writing the Yukawa couplings as $Y = {\rm diag}(y_{+},y_{-})$, we can express the dark sector Yukawa terms as
\begin{equation}
-\mathcal{L}_{y}=\overline{\Upsilon}_{L}(\Phi Y)\Upsilon_{R} + h.c. ,
\label{eq:yukmatrix}
\end{equation}
where
\begin{equation}
\Phi = (\, i \sigma^{2}\phi^{*} \, | \, \phi \,) \,\,\, .
\end{equation}
To simplify our subsequent analysis, we will assume $y_{+}= y_{-} \equiv y$, {\it i.e.}, that there is no isospin violation in the dark sector. 

Like QCD with two flavors, the dark sector has an $SU(2)_{L} \times SU(2)_{R}$ global chiral symmetry.   This symmetry is spontaneously broken to its diagonal subgroup (dark isospin) via dark fermion condensation
\begin{equation}
\langle p \, \overline{p} + m \, \overline{m} \rangle \approx 4 \pi  f^{3} \, ,
\end{equation}
where $f$ is the dark pion decay constant.   The low-energy effective theory can be described using the chiral Lagrangian approach where 
the triplet of dark pions $\Pi$ transform nonlinearly under the chiral symmetry, 
\begin{equation}
\Sigma = \exp(2\, i \, \Pi / f) \,\,\,\,\, ,  \,\,\,\,\, \Sigma \rightarrow L \, \Sigma \, R^\dagger \,\,\,,
\end{equation}
where $\Pi \equiv \pi^a \sigma^a/2$, and $\sigma^a$ are the Pauli matrices.   It is convenient to represent the fundamental scalar $\phi$ in
a similar way:
\begin{equation}
\Phi = \frac{\sigma+f'}{\sqrt{2}}\Sigma' \,\,\,\,\, , \,\,\,\,\, \Sigma'=\exp(2\, i\, \Pi' / f') \, .
\end{equation}
Recognizing that $\phi^\dagger \phi \equiv {\rm tr}(\Phi^{\dagger}\Phi)/2=(\sigma + f')^{2}/2$, we see that $\sigma$ 
parameterizes the fluctuation of the $\phi$ field about its vev, $f'$, in unitary gauge.  The kinetic terms for the scalar fields are given by 
\begin{align}
\mathcal{L}_{KE}   \, & =\frac{1}{2} \mbox{tr}(D_{\mu}\Phi^{\dagger} D^{\mu}\Phi)+\frac{f^{2}}{4}\mbox{tr}(D_{\mu}\Sigma^{\dagger} D^{\mu}\Sigma) \\
& =\frac{1}{2}\partial_{\mu}\sigma\partial^{\mu}\sigma+\frac{f^{2}}{4}\mbox{tr}(D_{\mu}\Sigma^{\dagger} D^{\mu}\Sigma)+\frac{(\sigma+f')^{2}}{4}\mbox{tr}(D_{\mu}\Sigma'^{\dagger} D^{\mu}\Sigma') \,\,\, .
\label{eq:chkin}
\end{align}
The SU(2)$_D$ covariant derivative $D_{\mu}=\partial_{\mu}-ig_{D}A^{a}_{\mu}\frac{\sigma^{a}}{2}$ that appears in Eq.~(\ref{eq:chkin}) will lead to terms involving a single derivative and a gauge field, allowing one to identify the unphysical scalar degrees of freedom that becomes the longitudinal components of the gauge fields.   Calling the unphysical and physics scalar degrees of freedom $\pi_u$ and $\pi_p$ respectively, one finds~\cite{Carone:2015jra}
\begin{align}
\pi_{u} &= \frac{f \, \Pi + f' \, \Pi'}{\sqrt{f^{2}+f'^{2}}} \\
\pi_{p} &= \frac{-f' \, \Pi +f \, \Pi'}{\sqrt{f^{2}+f'^{2}}}.
\end{align}
In addition to spontaneous breaking, the chiral symmetry is also explicitly broken by the Yukawa couplings in Eq.~(\ref{eq:yukmatrix}).   This effect may be captured in the chiral Lagrangian analysis by treating $\Phi Y$ as a ``spurion" with the following chiral transformation law
\begin{equation}
(\Phi Y) \rightarrow{} L(\Phi Y)R^{\dagger}.
\end{equation}
At lowest order, the term in the chiral Lagrangian involving the spurion is 
\begin{equation} 
\mathcal{L} = c_{1}4\pi f^{3}\mbox{tr}(\Phi Y \Sigma^\dagger) + h.c.,
\end{equation}. 
where the coefficient has been determined by naive dimensional analysis~\cite{NDA}, with the constant $c_{1}$ expected to be of order unity.   This term leads to a mass for the dark pion
\begin{equation}
m_\pi^2 = 2 c_{1}\sqrt{2}\frac{4\pi f}{f'}(f^2+f'^{2})\, y \,\,\, ,
\end{equation}
as well as a linear term for $\sigma$ that we take into account in our analysis of the scalar potential
\begin{equation}
V(\sigma)_{\rm lin} = -\kappa_{0}^{3} \, \sigma \,\,\, ,
\end{equation}
where $\kappa_{0}^{3} = 8\sqrt{2}\pi c_{1}f^{3}y$.   This term will provide the tilt in the scalar potential that leads to the electroweak-scale
vev of the Higgs field, after the tuning described below.

The quartic terms in the scalar potential are
\begin{equation}
V(\phi,H)_{\rm quart} = \frac{\lambda}{2}(H^{\dagger}H)^{2} - \lambda_{p}(H^{\dagger}H)(\phi^{\dagger}\phi) + \frac{\lambda_{\phi}}{2}(\phi^{\dagger}\phi)^{2}. \,\,\, ,
\end{equation}
where $H$ is the standard model Higgs doublet.   At large field amplitudes where these terms dominate, vacuum stability requires
\begin{equation}
\Delta =  \lambda\, \lambda_{\phi}  - \lambda_p^2 > 0 \,\,\,\, \text{ and } \,\,\,\,\, \lambda > 0 \,\, ,
\end{equation}
assuming $\lambda_p$ is non-vanishing. The two parameters, $\Delta$ and $\lambda$, define the positivity domain in parameter space; we will realize our tuning paradigm by taking $\Delta$ to be a very small, positive parameter.

After expanding the fields $H$ and $\phi$ in unitary gauge about their respective vevs, $v/\sqrt{2}$ and $f'/\sqrt{2}$, the tree-level potential may be written
\begin{equation}
V^{(0)}(\sigma,h) = \frac{\lambda}{8}(h+v)^4 - \frac{\lambda_{p}}{4}(h+v)^{2}(\sigma + f')^{2} + \frac{\lambda_{\phi}}{8}(\sigma + f')^{4} - \kappa_{0}^{3}\, \sigma .
\end{equation}
As in Ref.~\cite{Georgi:2024rxw}, we make the simplifying assumption that scalar mass terms are small enough to be omitted, since their presence only algebraically complicates the study of the tuning of interest.  Alternatively, one might say we have chosen to work with a ``classically conformal" model, one of a class of theories that have met considerable attention in the phenomenological literature. (See Ref.~\cite{Carone:2015jra} and references contained therein.)  Minimizing the potential allows us to identify the vacuum expectation values
\begin{equation}
f' = \kappa_0 \left(\frac{2 \lambda}{ \Delta} \right)^{1/3} \,\,\,\,\, \mbox{ and } \,\,\,\,\,
v=\kappa_0 \left(\frac{2 \lambda}{ \Delta} \right)^{1/3} \!\! \sqrt{\frac{\lambda_p}{ \lambda}}  \,\,\, .
\label{eq:thevevs}
\end{equation}
As one approaches the edge of the abyss, $\Delta \ll 1$, a hierarchy is created between the scales of the vevs and $\kappa_0$.   
The field-dependent mass squared matrix (which will be useful to us later) is given by
\begin{equation}
M^2(\sigma,h)=\begin{pmatrix}
\frac{3}{2} \, \lambda_{\phi} \, \sigma^{2}-\frac{1}{2} \, \lambda_{p} \, h^{2} & -  \lambda_{p} \, h \, \sigma \\
-  \lambda_{p} \, h \, \sigma& \frac{3}{2} \, \lambda\, h^{2}-\frac{1}{2} \,\lambda_{p} \,\sigma^{2}
\end{pmatrix} \,\, ,
\label{eq:fdmm}
\end{equation}
and yields the scalar mass matrix when evaluated at the vevs given in Eq.~(\ref{eq:thevevs}):
\begin{equation}
M^2 = 
\kappa_0^2 \left(\frac{2 \lambda}{ \Delta} \right)^{2/3} 
\begin{pmatrix}
(\lambda_p^2+\frac{3}{2} \Delta)/\lambda &\hspace{1em} &-\lambda_p^{3/2} / \lambda^{1/2}\\
-\lambda_p^{3/2} / \lambda^{1/2} & &\lambda_p 
\end{pmatrix} \,\,\, .
\label{eq:smsm}
\end{equation}
To parameterize the closeness to the edge of the positivity domain, we choose $\Delta$ as a free parameter 
in Eq.~(\ref{eq:smsm}) instead of $\lambda_{\phi}$, using $\lambda_{\phi} =(\Delta+\lambda_{p}^{2}) / \lambda$.
The eigenvalues of this matrix can be expressed exactly, but have particularly simple forms when written as an expansion
in $\Delta$.  Identifying the larger eigenvalues with the Higgs squared mass, we find
\begin{equation}
m_{h_0}^{2} = \frac{2^{2/3}\lambda_{p}\, (\lambda + \lambda_{p})}{\lambda^{1/3}} \frac{\kappa_{0}^{2}}{\Delta^{2/3}} + {\cal O}(\Delta^{1/3}) \,\,\, ,
\label{eq:mh0}
\end{equation} 
\begin{equation}
m_{\eta}^{2} = \frac{3}{2^{1/3}} \frac{\lambda^{2/3}}{(\lambda+\lambda_{p})}\, \kappa_{0}^{2} \, \Delta^{1/3}+ {\cal O}(\Delta^{4/3}) \,\,\, ,       
\label{eq:meta}
\end{equation}  
where we use $h_0$ and $\eta$ to refer to the mass eigenstate fields.  Unlike the vevs, one of the tree-level mass eigenvalues becomes
vanishingly small as $\Delta$ is tuned towards zero.  (As we will discuss later, this result is changed with the inclusion of loop corrections.)  The mixing angle that characterizes the two-dimensional rotation required to diagonalize Eq.~(\ref{eq:smsm}) also has
a simple form at lowest order in $\Delta$:
\begin{equation}
\tan^{2}{\theta} = \frac{\lambda_{p}}{\lambda}+ {\cal O}(\Delta) \,\,\,.
\label{eq:thetadef}
\end{equation}
As one might expect, the mixing angle vanishes as the portal coupling $\lambda_p$ approaches zero.

We now turn to the one-loop corrections, which have important physical effects.\footnote{Note that the one-loop corrections were not included in the analysis of Ref.~\cite{Carone:2015jra}.}  We first note that in the limit $\Delta \rightarrow 0$, the quartic terms terms
vanish along a ray in field space 
\begin{equation}
\sigma = \rho \cos\theta \,\,\,\,\,  \mbox{ and } \,\,\,\,\, h = \rho \sin\theta \,\,\, , 
\label{eq:ray}
\end{equation}
where $\rho$ is a parameter and the angle $\theta$ is also given by Eq.~(\ref{eq:thetadef}), or equivalently 
\begin{equation}
\cos\theta = \frac{\sqrt{\lambda}}{\sqrt{\lambda_p+\lambda}}  \,\,\, .
\end{equation}
One can check that a constrained minimization along this direction leads to a vev for $\rho$ that is consistent with the $\sigma$ and $h$ vevs shown in Eq.~(\ref{eq:thevevs}),
at lowest order in $\Delta$.    Graphically, the potential is very shallow along this ray and very steep perpendicular to it, which is also suggested by the eigenvalues of the mass squared matrix.   Due to the shallowness in this one direction, we expect one-loop corrections to be important (as they were in Ref.~\cite{Georgi:2024rxw}).    The general formula for the Coleman-Weinberg corrections~\cite{Coleman:1973jx} in a theory with a multi-field scalar sector can be found in Ref.~\cite{Chataignier:2018kay}: 
\begin{equation} 
V^{(1)}(\varphi_{i}) = \frac{1}{64\pi^{2}}\sum_{a}n_{a}M_{a}^{4}(\varphi_{i})\left(\log{\frac{M_{a}^{2}(\varphi_{i})}{\mu^{2}}}-C_{a}\right) \,\,\, ,    
\label{eq:cwgen}
\end{equation}
where the sum is over all particles, $M_a$ are the field-dependent masses, $\mu$ is the renormalization scale and the $\varphi_i$ represent the scalar fields; the constants $C_{a}=\frac{5}{6}$ for vector bosons 
and $C_{a}=\frac{3}{2}$ for all other types of particles.  If we let $s_{a}$ denote a particles spin, then $n_{a}$ is given by
\begin{equation}
n_{a} = (-1)^{2s_{a}}Q_{a}N_{a}(2s_{a}+1) \,\,\, ,
\label{eq:na}
\end{equation}
where $N_{a}=1$ ($N_{a}=3$) for uncolored (colored) particles, and $Q_{a}=1$ ($Q_{a}=2$ ) for electrically uncharged (charged) particles.  The $M_a^2$ and $n_a$ values are summarized in Table~\ref{table:1}. Note that the field dependent masses, $m_{h_0}^{2}(\sigma,h)$ and $m_{\eta}^{2}(\sigma,h)$ are the eigenvalues of the matrix shown in Eq.~(\ref{eq:fdmm}).

\begin{table}[h]
\centering
\begin{tabular}{cccccccccccc} \hline\hline
Species 	&&				&& $n_{a}$ && $M_{a}^{2}$ \\ \hline
 $h$ && &&  1 && $m_{h_0}^{2}(\sigma,h)$ \\
 $\eta$ && && 1 && $m_{\eta}^{2}(\sigma,h)$  \\
 $W^{\pm}$ && &&  6 && $g_{W}^{2}h^{2}/4$ \\
 $Z^{0}$ && &&  3 && $\sec^{2}\theta_{W}g_{W}^{2}h^{2}/4$ \\
 $t$ && &&  -12 && $y_{t}^{2}h^{2}/2$ \\
 $p$ && &&  -4 N &&  $y_{+}^{2}\sigma^{2}/2$  \\
 $m$ && && -4 N &&  $y_{-}^{2}\sigma^{2}/2$   \\
 $W_{D}$ && &&  9 && $g_{D}^{2}\sigma^{2}/4$ \\ \hline \hline
\end{tabular}
\caption{The field-dependent $M_a^2$ and the values of $n_a$ in Eq.~(\ref{eq:cwgen}) that define the Coleman-Weinberg corrections in the model.  Here $t$ and $y_t$ represent the top quark and its Yukawa coupling, respectively;  the $W^\pm$ and $Z$ are the electroweak gauge bosons, $\theta_{W}$ is the Weinberg angle, $g_{W}$ and $g_D$ are the SU(2)$_W$ and SU(2)$_D$ gauge couplings, respectively, and $N$ is the number of dark colors.}
\label{table:1}
\end{table}
In the spirit  of Ref.~\cite{Georgi:2024rxw}, we wish to choose the renormalization scale so that the vevs remain fixed at their tree-level values.   The complication is that there are two extremization conditions in the model but only one renormalization scale to fix.   We circumvent this difficulty by choosing the renormalization condition that the vev of $\rho$ 
remains fixed at its classical value:
\begin{equation} 
\log{\mu^{2}} = \frac{\sum\limits_a n_a M_a^2 \frac{\partial M_a^2}{\partial \rho} (\ln M_a^2-C_a+1/2)}{\sum\limits_a n_a M_a^2 \frac{\partial M_a^2}{\partial \rho}} \Bigg|_{\rho=\langle \rho \rangle} \,\,\, .    
\end{equation}
Of course, a different renormalization condition would change the meaning of the renormalized couplings, but would be no better or worse than the one above provided that the loop expansion remains perturbative for field values near the minimum of the potential.   For a generic field $\varphi$, the perturbativity condition is that the various $\alpha \ln (\varphi^2/\mu^2)$ remain small expansion parameters, where $\alpha$ is either a quartic coupling or the square of a gauge or Yukawa coupling divided by $4 \pi$~\cite{Sher:1988mj}.   In our numerical results presented in the next section, we have checked that there are no large logarithms near the potential minimum that would indicate a violation of the perturbative loop expansion.  

\begin{figure}[b]
 \includegraphics[width=0.48\textwidth]{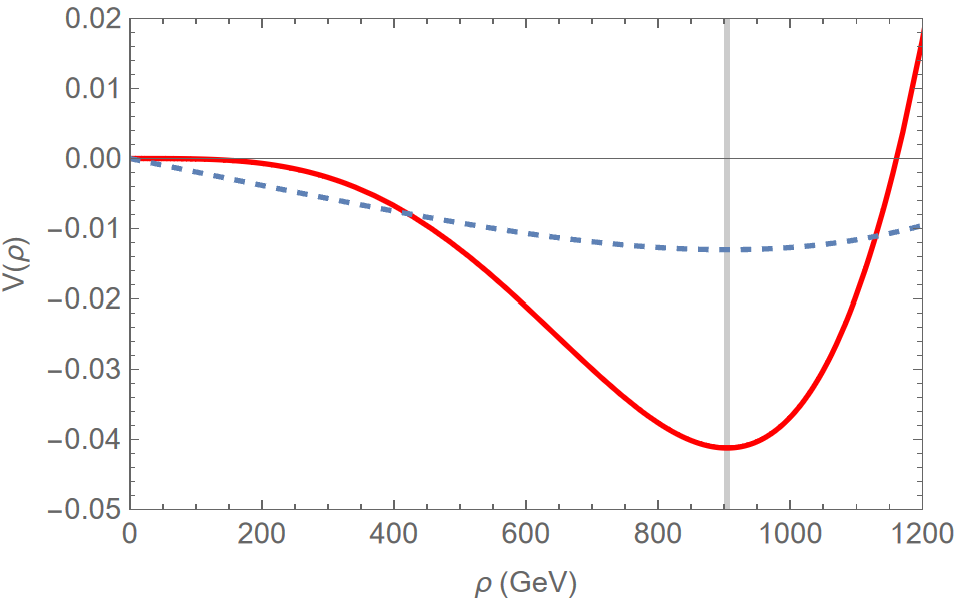}
            \caption{Comparison of the (nearly vanishing) tree-level potential and the one loop-corrected potential, where the latter has been multipled by $6 \times 10^{-9}$ to allow
            easier visual comparison (see the text for discussion).  The loop-corrected potential is represented by the curve with the deeper minimum.}
            \label{fig:one}
\end{figure}
In Fig.~\ref{fig:one}, we plot the potential along the ray defined by Eq.~(\ref{eq:ray}), to illustrate that our renormalization condition 
preserves the location of the minimum, which is deeper when the quantum corrections are included.  This behavior is qualitatively consistent with what was found in Ref.~\cite{Georgi:2024rxw}). The parameter values for the plot correspond to the example studied in the next section. Note that the difference in scale between the loop-corrected and tree-level results shown in the figure is to be expected since the ray was define by the condition that the dominant quartic terms in the tree-level potential vanish in this particular direction in field space.  As in Ref.~\cite{Georgi:2024rxw}, this does not reflect a breakdown of the loop expansion, but only the relative smallness of the tree-level terms in this particular direction in field space.  We comment more on the effect of the loop corrections in our discussion of the example presented in the next section.

\section{Dark Phenomenology}\label{sec:pheno}
In this section, we discuss some phenomenological aspects of the dark sector, to illustrate the separation of the scales involved, and also to show how a viable dark matter candidate can be obtained. 

Two quantities related to the Higgs sector are well known experimentally:  the Higgs boson mass, $125.2 \pm 0.11$~GeV~\cite{ParticleDataGroup:2024cfk}, and 
the Higgs field vev,  $v=246$~ GeV, where the latter sets the scale of the $W$ and $Z$ boson masses.   While we expect the one-loop corrections to be important in comparing 
the scalar sector predictions to data, it is nonetheless useful to look at the tree-level expressions first for some indication of preferred parameter ranges.  
Using the ratio of the tree-level expressions for $v$ and $m_{h_0}^2$   in Eqs.~(\ref{eq:thevevs})  and (\ref{eq:mh0}) respectively, one finds that
\begin{equation}
\lambda+\lambda_p \approx 0.259 \, .
\label{eq:lmlpsum}
\end{equation}
Mixing between the $h$ and $\sigma$ fields are described by a mixing angle $\theta$, where the relation to the mass eigenstates $h_0$ and $\eta$ 
are given by
\begin{equation}
\left(\begin{array}{c} h \\ \sigma \end{array}\right) = \left(\begin{array}{cc} \cos\theta & \sin\theta \\ -\sin\theta & \cos\theta \end{array}\right) \left(\begin{array}{c} h_0 \\ \eta \end{array}\right) \,\,\, .
\end{equation}
Mixing between the Higgs field and another neutral scalar field causes deviations from the standard model Higgs couplings. 
This leads to a lower bound on $\cos\theta$ that has been considered previously in the context of other Higgs portal dark matter models, 
$\cos\theta \geq 0.97$~\cite{Claude:2021sye}, or $\sin^2\theta < 0.0591$.  On the other hand, at lowest order in $\Delta$,
\begin{equation}
\sin^2\theta \approx \frac{\lambda_p}{\lambda+\lambda_p}  \,\, ,
\end{equation}
which together with Eq.~(\ref{eq:lmlpsum}) this implies
\begin{equation}
\lambda_p \alt 0.015 \,\,\,\,\, \mbox{ and } 0.244 \alt \lambda \alt 0.259 \,\,\, .
\label{eq:ranges}
\end{equation}
Given the need to obtain an adequate annihilation cross section for the dark matter in the model we will focus on values of $\lambda_p$ close to the upper bound given in Eq.~(\ref{eq:ranges}) and, hence, values of $\lambda$ closer to the lower limit of the range shown.  

There are two natural possibilities for dark matter in the model:  the dark pions $\pi_p$ and the dark baryons composed of the dark quarks 
$p$ and $m$.   The $\pi_p^\pm$ are exactly stable since they will be the lightest particles that are charged under the residual $Z_3$ symmetry 
described earlier that remains after spontaneous symmetry breaking. The neutral pions are also stable because (1) they have no decays to dark sector particles that are kinematically allowed, and (2) any decays to the visible sector particles would involve vertices of the form $\pi^0_p \, \sigma^n$, for $n \geq 1$, which are forbidden by the assumed isospin invariance of the dark sector and dark-visible sector interaction terms. Hence, we treat the whole dark pion triplet as stable dark  matter.  

We expect the dark baryon spectrum to have masses at or above the chiral symmetry breaking scale $4 \pi f$, which in the example below is significantly larger than $m_\pi$.   For $N>2$, the dark baryons are effectively stable due to an accidental dark baryon number symmetry, analogous to the baryon number symmetry in QCD.  If one assumes that the dark baryon-pion coupling is comparable to what is found
in QCD, $g_{\pi NN}^2 / (4 \pi) \sim 14$, then we estimate that annihilation of dark baryons and anti-baryons to $\pi_p \pi_p$ via the dark strong interactions, will lead to a negligible dark baryon contribution to the relic density for masses comparable to the ones discussed below.   However, there are other potential theoretical uncertainties in computing the dark baryonic component to the relic density which have led authors of similar models to omit consideration of the baryonic component~\cite{Hur:2011sv,Hur:2007uz}; for example there might be additional new physics that leads to matter-antimatter asymmetries in both the visible and dark sectors, which could affect the final result.  Exploring such possibilities go beyond the scope of the present discussion.   Alternatively, one might eliminate the stabilizing baryon number symmetry by restricting to the case of $N=2$ dark colors,  where dark baryon number can be violated at tree-level by $Z_3$-invariant mass-mixing terms of the form $\overline{p_R^c} m_R$, with the superscript $c$ representing charge conjugation.  In the case where such terms are introduced as small perturbations, the stabilizing baryon number symmetry is eliminated but the approximate chiral symmetry structure of the theory as we have described it remains intact, with the lighter dark pion states providing the dark matter.

While $\pi_p$ dark matter was considered in Ref.~\cite{Carone:2015jra}, the present scenario is different since the dark matter is now much lighter than the TeV-scale.  The  $\pi_p \pi_p \rightarrow \eta \, \eta$ annihilation channel considered in that work, which scales as roughly the square of this scale, is too 
small for this process to efficiently annihilate away enough dark matter assuming the thermal freeze-out mechanism.  Here, we can obtain the correct 
relic density by resonant annihilation
\begin{equation}
\pi_p \pi_p \rightarrow f \, \overline{f} \,\,\, ,
\end{equation}
where we have an $\eta$ exchanged in the $s$-channel and $m_\eta$ is near $2 \, m_\pi$.  For the purposes of numerical estimate, we need the $\eta$ width 
$\Gamma \equiv \sum_f \Gamma(\eta\rightarrow f \, \overline{f})$, where
\begin{equation}
\Gamma(\eta\rightarrow f \, \overline{f}) = \frac{N^f_c}{8\pi} \frac{m_f^2}{v^2} \sin^2 \!\theta\, m_\eta \left(1-\frac{4 m_f^2}{m_\eta^2}\right)^{3/2} \,\,\, ,
\end{equation}
where $m_f$ is a standard model fermion mass, and $N^f_c=3$ ($1$) for quarks (leptons).  We find that the nonrelativistic limit of resonant annihilation cross section times relative velocity
\begin{equation}
v_{\rm rel} \, \sigma_{\rm ann} = \sum_f \frac{N^f_c}{16 \pi}\frac{\sin^2 \!2 \theta \, m_f^2}{f'^2 \, v^2}\frac{m_\pi \, (m_\pi^2-m_f^2)^{3/2}}{(4 \, m_\pi^2-m_\eta^2)^2+m_\eta^2\,  \Gamma^2} \,\,\,,
\end{equation}
where the sum extends over all $f$ such that $m_f \leq m_\pi$.   The freeze-out temperature $T_F$  is determined by the point where 
\begin{equation}
n^{\rm EQ}_\pi \langle \sigma v_{\rm rel} \rangle / H(T_F) \approx 1 \,\, , 
\end{equation}
where $n^{\rm EQ}$ is the nonrelativistic equilibrium number density, and $H$ is the Hubble parameter for a radiation dominated universe.  Expressions for both may be found in standard texts~\cite{kandt}.
Defining $x_F=m_\pi/T_f$, the relic density 
\begin{equation}
\Omega_D h^2 \approx \frac{3 \, (1.07 \times 10^9\mbox{ GeV}^{-1}) \, x_F}{\sqrt{g_*(x_F)} \, M_{\rm Pl} \langle \sigma v_{\rm rel} \rangle_F} \,\,\, .
\end{equation}

A benchmark point that produces the correct Higgs mass and vev has the following input parameters:
\begin{equation} 
\begin{array}{ccccccc}
\Delta = 9.0 \times 10^{-13}, & & \lambda = 0.1650, & \hspace{0.5 em} & \lambda_{p} = 1.318 \times 10^{-2}, &\hspace{1em} & \kappa_{0} = 0.1216 \mbox{ GeV}^3,\\
g_{D} = 0.49158, & \hspace{1 em}& y = 7.3333 \times 10^{-5}, & & c_1=1.0 ,  & & N=4 \, . \\      
\end{array}    
\label{eq:inparams}
\end{equation}
Using the one-loop corrected potential to determine the scalar mass eigenvalues, the remaining output parameters for this choice are:
\begin{equation} 
\begin{array}{ccccccc}
f=0.8836\mbox{ GeV}, &\hspace{1em} & f'=870.3\mbox{ GeV}, & \hspace{1em} &m_\eta \approx  2.831\mbox{ GeV},&\hspace{1em} & m_\pi \approx 1.416\mbox{ GeV} ,\\
\sin^2\theta=0.034, & & \lambda_\phi=1.053 \times 10^{-3} ,&  & m_{W_D}= 213.9\mbox{ GeV}, & & m_\pm= 45.13\mbox{ MeV} \, . \\    
\end{array}   
\label{eq:outparams}
\end{equation}
This example illustrates the main features of the dark abyss scenario.   The value of $\Delta \ll 1$ parameterizes the tuning towards an edge of the positivity domain of the potential.   With 
the dynamical scale of the dark sector fixed through the value of $\kappa_0$ (which, after specifying the input values of $c_1$ and $y$, determines the dark pion decay constant $f$), 
the tuning allows for the much higher scale of the vev of the $\sigma$ field $f'$, which in turn triggers a somewhat lower value of the weak scale $v \approx 246$~GeV via mixing effects.   
The value of $\lambda_p$ is consistent with Eq.~(\ref{eq:ranges}), where $\lambda$ is below the tree level range, as a consequence of the one-loop corrections.   The loop corrections have 
a more significant effect on the lighter scalar mass eigenvalue, $m _\eta$, which can be understood by noting that the tree-level contribution Eq.~(\ref{eq:meta}) approaches zero as 
$\Delta \rightarrow 0$.  The exact value of $m_\eta$ depends on the relative size of the positive and negative contributions of the loop corrections to the second-derivative matrix of the 
potential, where the sign differences originate from Eq.~(\ref{eq:na}).   Since the value of the SU(2)$_D$ gauge coupling is unknown and controls one of the positive contributions, we can 
choose it to adjust the value of $m_\eta$, to obtain the example presented above.  As discussed earlier, we have omitted scalar  mass terms to simplify the analysis~\cite{Georgi:2024rxw}; allowing them would provide additional freedom to adjust the scalar mass eigenvalues.  For this example,  we find numerically that  $\Omega_D h^2 \approx 0.12$, which was by design: the pion and $\eta$ masses were chosen to be as close as needed so that resonant annihilation would provide the correct relic density.   We present this only as an existence proof; the dark abyss scenario in which a higher electroweak scale is triggered from a much lighter dynamical scale in a dark sector should be operative in models where the dark matter phenomenology is entirely different.   For example, one might upgrade the $Z_3$ symmetry to a U(1) gauge symmetry, so that a dark photon can contribute to dark matter annihilation;  alternatively, one might consider models without imposing an exact $Z_3$ to allow the dark pions to decay, while dark matter is provided by another source (for example, QCD axions).  Surveying the full parameter space on any one model is not the purpose of this paper, but we hope the example presented in this section illustrates the basic features of the framework of interest.

Finally, we note that dark matter-nucleon elastic scattering in this model occurs through $h_0$ and $\eta$ exchange, as in Ref.~\cite{Carone:2015jra}, so we evaluate the spin-independent scattering cross section formula presented there for the parameters in Eqs.~(\ref{eq:inparams}) and (\ref{eq:outparams}).   We find $\sigma_{SI} \approx 2.4 \times 10^{-44}$~cm$^2$.   The current bound for dark matter with mass around 1 GeV is ${\cal O}(10^{-42})$ cm$^2$ from the DarkSide-50 experiment~\cite{DarkSide:2022dhx}, indicating that our benchmark point is not excluded by current bounds.  

\section{Conclusions}\label{sec:conc}
Naturalness has provided a guiding principle for a wide range of beyond-the-standard-model theories, whose additional particles and interactions cancel the quadratic divergences
of the standard model.   While such theories reduce the sensitivity of the weak scale to physics at much higher scales, the new physics predicted by these theories at 
potentially observable energies has, thus far,  eluded all experimental searches.  In this paper, we have set aside the goal of eliminating fine tuning and considered instead the
implications of a particular tuning paradigm proposed in Ref.~\cite{Georgi:2024rxw}.   This approach restricts the parameter space of the scalar potential to be very near the boundary 
between global vacuum stability and instability.  A cubic term in the potential with a small dimensionful coefficient can drive the field vacuum expectation values (vevs) to much higher 
scales than those present in the Lagrangian.  In this paper, we explored how this type of tuning can achieve a large separation between the scale of a light dark sector and the electroweak scale.  One may think of cubic terms as being induced after shifting away a linear term in the scalar potential that arises when fermions condense due to dark sector strong 
dynamics.   A single generation of dark-colored fermions has a global chiral symmetry that is spontaneously broken to dark isospin; the relevant physics can be described using a chiral Lagrangian approach that is familiar from QCD or similar technicolor models~\cite{Carone:1993xc}. The dark sector includes a weakly coupled SU(2)$_D$ gauge group that is spontaneously broken by a 
doublet $\phi$ that couples to the dark fermions and whose vacuum expectation value (vev) is triggered by their condensate.   Due to tuning towards the edge of the abyss, the doublet 
vev is much larger than the scale of the condensate, and its coupling to the standard model Higgs doublet generates the electroweak scale via mixing effects.\footnote{One might wonder whether the electroweak scale could be generated directly from the QCD condensate via a similar mechanism without new physics, by tuning the Higgs quartic coupling towards small values.  One finds in this case that the Coleman-Weinberg corrections change the classical minimum to a maximum, preventing a viable solution.}     

For the purpose of illustration, we study this model numerically at a benchmark point in parameter space.   We show that the parameter space is substantially constrained in the given tuning paradigm by the requirements that the measured Higgs boson mass and vev are reproduced, and that the mixing angle in the scalar sector remains below its experimental bound.  We also show that the inclusion of one-loop corrections to the potential deepens the desired minimum of the potential compared to the tree-level result, in agreement with the behavior noted in Ref.~\cite{Georgi:2024rxw}.   The dark pion is stable due to a discrete symmetry in the theory and can serve as a dark matter candidate.   The example presented achieves the correct dark matter relic density through a resonant annihilation process, while remaining consistent with the experimental bound on the dark matter-nucleon spin independent scattering cross section.  We give a few examples of other possible dark sectors that may trigger the electroweak scale in a similar manner.

While the current work neither exhaustively studies the parameter space of a single model (one that is not likely to be the ultimate theory of nature), or survey a wide variety of similar models, we hope that it has illustrated the utility of this tuning paradigm as an organizing principle for model building via an example that differs from the original proposal.   It is our hope that applications of this tuning in other beyond-the-standard-model settings in which there are separations of scales (for example, in flavor models) may lead to theoretical insights and to interesting phenomenology. 
\begin{acknowledgments} 
We thank the NSF for support under Grant No. PHY-2112460 and No. PHY-2411549.
\end{acknowledgments}



\begin{thebibliography}{99}

\bibitem{Martin:1997ns}
See, for example, S.~P.~Martin,
``A Supersymmetry primer,''
Adv. Ser. Direct. High Energy Phys. \textbf{18}, 1-98 (1998),
\href{https://arxiv.org/abs/hep-ph/9709356}{arXiv:hep-ph/9709356 [hep-ph]}.

\bibitem{Perelstein:2005ka}
M.~Perelstein,
``Little Higgs models and their phenomenology,''
Prog. Part. Nucl. Phys. \textbf{58}, 247-291 (2007),
\href{https://arxiv.org/abs/hep-ph/0512128}{arXiv:hep-ph/0512128 [hep-ph]}.

\bibitem{Grinstein:2007mp}
B.~Grinstein, D.~O'Connell and M.~B.~Wise,
``The Lee-Wick standard model,''
Phys. Rev. D \textbf{77}, 025012 (2008),
\href{https://arxiv.org/abs/0704.1845}{arXiv:0704.1845 [hep-ph]}.

\bibitem{Carone:2008iw}
C.~D.~Carone and R.~F.~Lebed,
``A Higher-Derivative Lee-Wick Standard Model,''
JHEP \textbf{01}, 043 (2009),
\href{https://arxiv.org/abs/0811.4150}{arXiv:0811.4150 [hep-ph]}.

\bibitem{PDGNavas}
See, for example, ``Supersymmetry, Part II (Experiment)" in 
Ref.~\cite{ParticleDataGroup:2024cfk}

\bibitem{Georgi:2024rxw}
H.~Georgi,
``Tuning to the edge of the abyss in SU(5),''
Phys. Lett. B \textbf{853}, 138703 (2024),
\href{https://arxiv.org/abs/2402.09331}{arXiv:2402.09331 [hep-ph]}.

\bibitem{Froggatt:1995rt}
C.~D.~Froggatt and H.~B.~Nielsen,
``Standard model criticality prediction: Top mass $173 \pm 5$ GeV and Higgs mass $135 \pm 9$ GeV,''
Phys. Lett. B \textbf{368}, 96-102 (1996),
\href{https://arxiv.org/abs/hep-ph/9511371}{arXiv:hep-ph/9511371 [hep-ph]}.

\bibitem{criticality}
C.~Er\"oncel, J.~Hubisz and G.~Rigo,
``Self-Organized Higgs Criticality,''
JHEP \textbf{03}, 046 (2019),
\href{https://arxiv.org/abs/1804.00004}{arXiv:1804.00004 [hep-ph]};
J.~Khoury, ``Accessibility Measure for Eternal Inflation: Dynamical Criticality and Higgs Metastability,''
JCAP \textbf{06}, 009 (2021),
\href{https://arxiv.org/abs/1912.06706}{arXiv:1912.06706 [hep-th]};
G.~Kartvelishvili, J.~Khoury and A.~Sharma,
``The Self-Organized Critical Multiverse,''
JCAP \textbf{02}, 028 (2021),
\href{https://arxiv.org/abs/2003.12594}{arXiv:2003.12594 [hep-th]};
G.~F.~Giudice, M.~McCullough and T.~You,
``Self-organised localisation,''
JHEP \textbf{10}, 093 (2021),
\href{https://arxiv.org/abs/2105.08617}{arXiv:2105.08617 [hep-ph]}.

\bibitem{Bak:1987xua}
P.~Bak, C.~Tang and K.~Wiesenfeld,
``Self-organized criticality: An Explanation of 1/f noise,''
Phys. Rev. Lett. \textbf{59}, 381-384 (1987).

\bibitem{Isidori:2001bm}
For example, see G.~Isidori, G.~Ridolfi and A.~Strumia,
``On the metastability of the standard model vacuum,''
Nucl. Phys. B \textbf{609}, 387-409 (2001),
\href{https://arxiv.org/abs/hep-ph/0104016}{arXiv:hep-ph/0104016 [hep-ph]}.

\bibitem{ParticleDataGroup:2024cfk}
S.~Navas \textit{et al.} [Particle Data Group],
``Review of particle physics,''
Phys. Rev. D \textbf{110}, no.3, 030001 (2024)

\bibitem{Claude:2021sye}
J.~Claude and S.~Godfrey,
``Exploring Direct Detection Suppressed Regions in a Simple 2-Scalar Mediator Model of Scalar Dark Matter,''
Eur. Phys. J. C \textbf{81}, no.~5, 405 (2021),
\href{https://arxiv.org/abs/2104.01096}{[arXiv:2104.01096 [hep-ph]}.

\bibitem{Carone:2015jra}
C.~D.~Carone and R.~Ramos,
``Dark chiral symmetry breaking and the origin of the electroweak scale,''
Phys. Lett. B \textbf{746}, 424-429 (2015),
\href{https://arxiv.org/abs/1505.04448}{arXiv:1505.04448 [hep-ph]}.

\bibitem{Witten:1982fp}
E.~Witten,
``An SU(2) Anomaly,''
Phys. Lett. B \textbf{117}, 324-328 (1982).

\bibitem{Carone:1993xc}
C.~D.~Carone and H.~Georgi,
``Technicolor with a massless scalar doublet,''
Phys. Rev. D \textbf{49}, 1427-1436 (1994),
\href{https://arxiv.org/abs/hep-ph/9308205}{[arXiv:hep-ph/9308205 [hep-ph]}.

\bibitem{Ibanez:1991hv}
L.~E.~Ibanez and G.~G.~Ross,
``Discrete gauge symmetry anomalies,''
Phys. Lett. B \textbf{260}, 291-295 (1991).

\bibitem{Banks:1991xj}
T.~Banks and M.~Dine,
``Note on discrete gauge anomalies,''
Phys. Rev. D \textbf{45}, 1424-1427 (1992),
\href{https://arxiv.org/abs/hep-th/9109045}{arXiv:hep-th/9109045 [hep-th]}.

\bibitem{NDA}
A.~Manohar and H.~Georgi,
``Chiral Quarks and the Nonrelativistic Quark Model,''
Nucl. Phys. B \textbf{234}, 189-212 (1984);
H.~Georgi and L.~Randall,
``Flavor Conserving CP Violation in Invisible Axion Models,''
Nucl. Phys. B \textbf{276}, 241-252 (1986).

\bibitem{Coleman:1973jx}
S.~R.~Coleman and E.~J.~Weinberg,
``Radiative Corrections as the Origin of Spontaneous Symmetry Breaking,''
Phys. Rev. D \textbf{7}, 1888-1910 (1973).

\bibitem{Chataignier:2018kay}
L.~Chataignier, T.~Prokopec, M.~G.~Schmidt and B.~\'Swie\.zewska,
``Systematic analysis of radiative symmetry breaking in models with extended scalar sector,''
JHEP \textbf{08}, 083 (2018),
\href{https://arxiv.org/abs/1805.09292}{arXiv:1805.09292 [hep-ph]}.

\bibitem{Sher:1988mj}
M.~Sher,
``Electroweak Higgs Potentials and Vacuum Stability,''
Phys. Rept. \textbf{179}, 273-418 (1989).

\bibitem{Hur:2011sv} 
T.~Hur and P.~Ko,
``Scale invariant extension of the standard model with strongly interacting hidden sector,''
Phys. Rev. Lett. \textbf{106}, 141802 (2011),
\href{https://arxiv.org/abs/1103.2571}{arXiv:1103.2571 [hep-ph]}.

\bibitem{Hur:2007uz}
T.~Hur, D.~W.~Jung, P.~Ko and J.~Y.~Lee,
``Electroweak symmetry breaking and cold dark matter from strongly interacting hidden sector,''
Phys. Lett. B \textbf{696}, 262-265 (2011),
\href{https://arxiv.org/abs/0709.1218}{arXiv:0709.1218 [hep-ph]}.

\bibitem{kandt}
Kolb, E. and Turner, M., {\it The Early Universe} (Frontiers in Physics), Addison and Wesley, Redwood City, California, USA, 1990.

\bibitem{DarkSide:2022dhx}
P.~Agnes \textit{et al.} [DarkSide],
``Search for Dark-Matter\textendash{}Nucleon Interactions via Migdal Effect with DarkSide-50,''
Phys. Rev. Lett. \textbf{130}, no.~10, 101001 (2023),
\href{https://arxiv.org/abs/2207.11967}{arXiv:2207.11967 [hep-ex]}.

\end{thebibliography}
\end{document}